\documentclass[preprint,nofootinbib]{revtex4}%
\usepackage{amssymb}
\usepackage{amsfonts}
\usepackage{amsmath}
\usepackage{graphicx}%
\newcommand{\beq}{\begin{equation}}
\newcommand{\eeq}{\end{equation}}
\newcommand{\bea}{\begin{eqnarray}}
\newcommand{\eea}{\end{eqnarray}}
\begin{document}
\title{More on Asymptotically Anti-de Sitter Spaces\\in Topologically Massive Gravity}
\author{Marc Henneaux$^{1,2}$, Cristi\'{a}n Mart\'{\i}nez$^{1,3}$, Ricardo
Troncoso$^{1,3}$}
\email{henneaux@ulb.ac.be, martinez@cecs.cl, troncoso@cecs.cl}
\affiliation{$^{1}$Centro de Estudios Cient\'{\i}ficos (CECS), Casilla 1469, Valdivia, }
\affiliation{$^{2}$Physique th\'{e}orique et math\'{e}matique, Universit\'{e} Libre de
Bruxelles and International Solvay Institutes,ULB Campus Plaine C.P.231,
B-1050 Bruxelles, Belgium, }
\affiliation{$^{3}$Centro de Ingenier\'{\i}a de la Innovaci\'{o}n del CECS (CIN), Valdivia, Chile.}
\preprint{CECS-PHY-10/07}

\begin{abstract}
Recently, the asymptotic behaviour of three-dimensional anti-de Sitter gravity
with a topological mass term was investigated. Boundary conditions were given
that were asymptotically invariant under the two-dimensional conformal group
and that included a fall-off of the metric sufficiently slow to consistently
allow pp-wave type of solutions. Now, pp-waves can have two different
chiralities. Above the chiral point and at the chiral point, however, only one
chirality can be considered, namely the chirality that has the milder
behaviour at infinity. The other chirality blows up faster than AdS and does
not define an asymptotically AdS spacetime. By contrast, both chiralities are
subdominant with respect to the asymptotic behaviour of AdS spacetime below
the chiral point. Nevertheless, the boundary conditions given in the earlier
treatment only included one of the two chiralities (which could be either one)
at a time. We investigate in this paper whether one can generalize these
boundary conditions in order to consider simultaneously both chiralities below
the chiral point. We show that this is not possible if one wants to keep the
two-dimensional conformal group as asymptotic symmetry group. Hence, the
boundary conditions given in the earlier treatment appear to be the best
possible ones compatible with conformal symmetry. In the course of our
investigations, we provide general formulas controlling the asymptotic charges
for all values of the topological mass (not just below the chiral point).


\end{abstract}
\maketitle



\section{Introduction}

\setcounter{equation}{0} Topologically massive gravity in three dimensions
with a negative cosmological constant \cite{Deser:1982vy,DeserJT,DeserCosmo},
described by the action
\begin{align}
I[e]  &  =2\int\left[  e^{a}\left(  d\omega_{a}+\frac{1}{2}\epsilon
_{abc}\omega^{b}\omega^{c}\right)  +\frac{1}{6}\frac{1}{\ell^{2}}%
\epsilon_{abc}e^{a}e^{b}e^{c}\right] \nonumber\\
&  +\frac{1}{\mu}\int\left[  \omega^{a}\left(  d\omega_{a}+\frac{1}{3}%
\epsilon_{abc}\omega^{b}\omega^{c}\right)  \right]  \label{actionTMG}%
\end{align}
admits a rich variety of non trivial solutions (see
\cite{Chow:2009km,Chow:2009vt} for recent reviews and new
solutions)\footnote{Our conventions are as follows: $\mu\neq0$ is the mass
parameter, $\ell$ is the AdS radius, $\epsilon_{012}=1$ (so $\epsilon^{{012}%
}=-1$) and we have set the gravitational coupling constant $16\pi G=1$.}.
Among these, pp-waves \cite{DS, OST} are particularly interesting as they
preserve supersymmetry \cite{Gibbons:2008vi}. For a given value of the
topological mass parameter $\mu\ell\not =\pm1$, there are two chiralities,
described by
\begin{equation}
ds^{2}=\ell^{2}\frac{dr^{2}}{r^{2}}-r^{2}dx^{+}dx^{-}+F(x^{-})r^{1-\mu\ell
}\left(  dx^{-}\right)  ^{2} \label{pp-wave-mu-}%
\end{equation}
(negative chirality) and
\begin{equation}
ds^{2}=\ell^{2}\frac{dr^{2}}{r^{2}}-r^{2}dx^{+}dx^{-}+G(x^{+})r^{1+\mu\ell
}\left(  dx^{+}\right)  ^{2} \label{pp-wave-mu+}%
\end{equation}
(positive chirality), where $F$ and $G$ are arbitrary functions. We assume the
basis $\{\frac{\partial}{\partial r},\frac{\partial}{\partial x^{+}}%
,\frac{\partial}{\partial x^{-}}\}$ to have positive orientation.

For a standard pp-wave, the coordinates $t$ and $\phi$, related to $x^{+}$
and $x^{-}$ according to $x^{\pm}=\frac{t}{\ell}\pm\phi$, are assumed to be of
infinite range, $-\infty<t<+\infty$, $-\infty<\phi<+\infty$. However, one may
clearly regard $\phi$ as an angle, $0\leq\phi\leq2\pi$ without changing the
fact that (\ref{pp-wave-mu-}) and (\ref{pp-wave-mu+}) are solutions. This is
what we shall do here since we want to study spaces which are asymptotical to
anti-de Sitter space whose metric reads in standard static coordinates
\[
d\bar{s}^{2}\!=\!\!\displaystyle\left(  \!1+\displaystyle\frac{r^{2}}{\ell
^{2}}\right)  ^{-1}\!\!\!\!\!\!dr^{2}-\frac{\ell^{2}}{4}(dx^{+2}%
\!+dx^{-2})-\left(  \!\frac{\ell^{2}}{2}+r^{2}\!\!\right)  \!dx^{+}dx^{-},
\]
where $\phi$ is an angle.

In the chiral case $\mu\ell=\pm1$, which has attracted much attention recently
\cite{Attention} following the lead of \cite{LSStrominger}, the pp-wave
solutions acquire a logarithmic behaviour in $r$. {}For $\mu\ell=1$, they read
explicitly
\begin{equation}
ds^{2}=\ell^{2}\frac{dr^{2}}{r^{2}}-r^{2}dx^{+}dx^{-}+F(x^{-})\log r\left(
dx^{-}\right)  ^{2} \label{pp-wave-mu-chiral}%
\end{equation}
(negative chirality) and
\begin{equation}
ds^{2}=\ell^{2}\frac{dr^{2}}{r^{2}}-r^{2}dx^{+}dx^{-}+G(x^{+})\,r^{2}\,\log
r\left(  dx^{+}\right)  ^{2} \label{pp-wave-mu+chiral}%
\end{equation}
(positive chirality), where $F$ and $G$ are again arbitrary functions. The
solutions for $\mu\ell=-1$ are simply obtained by exchanging $x^{+}$ and
$x^{-}$.

When $\vert\mu\ell\vert\geq1$, only one of the solutions (\ref{pp-wave-mu-})
and (\ref{pp-wave-mu+}) (or (\ref{pp-wave-mu-chiral}) and
(\ref{pp-wave-mu+chiral})) asymptotically matches AdS in the given coordinate
system, namely the negative chirality solution if $\mu\ell\geq1$ or the
positive chirality solution if $\mu\ell\leq-1$. Indeed, it is only for that
solution that the term containing $F$ or $G$ is subdominant with respect to
the asymptotic behaviour $\sim r^{2}$ of the angular part of the AdS metric.
For the other solution the $g_{++}$ ($g_{--}$) component goes to infinity
faster than $r^{2}$. When $\vert\mu\ell\vert< 1$, however, both solutions are
such that $g_{++}$ and $g_{--}$ blow up at infinity more slowly than $r^{2}$.

In \cite{Henneaux:2009pw}, boundary conditions are devised satisfying the
consistency requirements spelled out in \cite{Henneaux:1985tv}:

\begin{itemize}
\item They are invariant under the anti-de Sitter group.

\item They decay sufficiently slowly to the exact anti-de Sitter metric at
infinity so as to contain the \textquotedblleft asymptotically anti-de Sitter"
solutions of the theory of physical interest (see below for precise statements
on this point).

\item But at the same time, the fall-off is sufficiently fast so as to yield
finite charges.
\end{itemize}

The boundary conditions of \cite{Henneaux:2009pw} generalize those of
\cite{Brown:1986nw} by accommodating more solutions. Not only do they contain
the BTZ black holes \cite{BTZ,BTZHenneaux}, as \cite{Brown:1986nw} does, but
they also include in addition the slower asymptotic behaviour of the above
pp-wave metrics. More precisely, they enable one to turn on one chirality,
which is the chirality that does not blow up faster than AdS when $|\mu
\ell|\geq1$, and which can be any of the two chiralities when $|\mu\ell|<1$.

In that latter case, however, the boundary conditions of
\cite{Henneaux:2009pw} only allow to switch on one chirality at a time. This
appears to be sufficient to accommodate the known exact solutions, which do
not involve both chiralities. It is nevertheless somewhat disturbing since
each chirality is then individually compatible with the asymptotic anti-de
Sitter symmetry. One might in principle, by integrating from infinity,
construct solutions where both chiralities are present.

It is therefore natural to ask whether one can extend the boundary conditions
of \cite{Henneaux:2009pw} so as to allow simultaneously both chiralities when
$|\mu\ell|<1$.

The situation is somewhat similar to what happens when a massive scalar field
is present, which has been extensively studied in
\cite{Henneaux:2002wm,Henneaux:2004zi,Henneaux:2006hk,Hertog-Maeda,Marolf}.
When the mass $m$ is in the range $m_{BF}^{2}<m^{2}<m_{BF}^{2}+1/l^{2}$ (where
$m_{BF}$ is the Breitenlohner-Freedman bound \cite{B-F}), there are two
possible admissible behaviours (two \textquotedblleft branches") for the
scalar field. One may device boundary conditions which exclude one of the two
branches. But one can also consider more general boundary conditions where
both branches are switched on while preserving the asymptotically anti-de
Sitter group.

We show in this paper that similarly a more general asymptotic treatment of
topologically massive gravity exists when $|\mu\ell|<1$. This more general
treatment allows both chiralities. However, contrary to what happens in the
scalar field case, asymptotic anti-de Sitter invariance is lost when both
chiralities are simultaneously switched on\footnote{The results of this
analysis were announced (but not proved) in \cite{Henneaux:2009pw}.}. It
appears therefore that the boundary conditions of \cite{Henneaux:2009pw} are
the most general boundary conditions compatible with full anti-de Sitter symmetry.

What happens can be roughly understood as follows (the precise analysis is
given below). The two chiralities are conjugate, as are the two branches of
the scalar field. Hence, in the variational principle, one cannot vary them
independently but one must fix a relation between them, just as one cannot
leave both the q's and the p's free in the Hamiltonian variational principle.
While in the scalar case, the relation between both branches can be chosen to
be anti-de Sitter invariant without killing one branch, this is not the case
here. The only anti-de Sitter invariant relations are obtained by setting one
of the two chiralities to zero.

Our paper is organized as follows. In the next section, we give the boundary
conditions that incorporate both chiralities. When one sets one chirality to
zero, one recovers the boundary conditions of \cite{Henneaux:2009pw}. We then
show that these boundary conditions, without additional restrictions, are
invariant under the full conformal group at infinity. However, the need to
have well-defined charges, whose variations are not only finite but also
integrable, forces one to impose a relation on the two chiralities. In Section
\ref{Surface Terms}, we prove that there is no such relation that preserves
the conformal symmetry at infinity, except setting one of the two chiralities
to zero. The best that one can achieve otherwise is Virasoro$\times R$.

We collect in the appendices the technical tools necessary for handling the
surface integrals. This gives us the opportunity to provide the detailed
formulas promised in \cite{Henneaux:2009pw} (a special appendix is in
particular reserved to the chiral point).

\section{Boundary conditions for TMG with $\vert\mu\ell\vert< 1$}

\setcounter{equation}{0}

\subsection{Asymptotic Conditions}

We tentatively take as boundary conditions
\begin{equation}%
\begin{array}
[c]{lll}%
\Delta g_{rr} & = & \!f_{rr}r^{-4}+\cdot\cdot\cdot\\
\Delta g_{r+} & = & \!h_{r+}\ r^{-2+\mu l}+\!f_{r+}r^{-3}+\cdot\cdot\cdot\\
\Delta g_{r-} & = & \!h_{r-}\ r^{-2-\mu l}+f_{r-}r^{-3}+\cdot\cdot\cdot\\
\Delta g_{++} & = & \!h_{++}\ r^{1+\mu l}+\!f_{++}+\cdot\cdot\cdot\\
\Delta g_{+-} & = & \!f_{+-}+\cdot\cdot\cdot\\
\Delta g_{--} & = & \!h_{--}\ r^{1-\mu l}+f_{--}+\cdot\cdot\cdot
\end{array}
\label{Asympt relaxed metric mu}%
\end{equation}
where $f_{\mu\nu}$ and $h_{\mu\nu}$ depend only on $x^{+}$ and $x^{-}$ and not
on $r$. We use the convention that the $f$-terms are the standard deviations
from AdS considered in \cite{Brown:1986nw}, while the $h$-terms represent the
relaxed terms that need to be included in order to accommodate the solutions
of the topologically massive theory with slower fall-off.

When only one chirality is included, one recovers the boundary conditions of
\cite{Henneaux:2009pw}. This is how (\ref{Asympt relaxed metric mu}) was
arrived at, by simply superposing the boundary conditions for the individual
chiralities considered in that reference.

We shall see below that these boundary conditions need to be strengthened.
Integrability of the charges forces indeed a relationship between $h_{++}$ and
$h_{--}$.

\subsection{Asymptotic Symmetry}

One easily verifies that the asymptotic conditions without the extra relation
$h_{++}= h_{++}(h_{--})$ are invariant under diffeomorphisms that behave at
infinity as
\begin{align}
\eta^{+}  &  =T^{+}+\frac{l^{2}}{2r^{2}}\partial_{-}^{2}T^{-}+\cdot\cdot
\cdot\nonumber\\
\eta^{-}  &  =T^{-}+\frac{l^{2}}{2r^{2}}\partial_{+}^{2}T^{+}+\cdot\cdot
\cdot\label{Asympt KV}\\
\eta^{r}  &  =-\frac{r}{2}\left(  \partial_{+}T^{+}+\partial_{-}T^{-}\right)
+\cdot\cdot\cdot\nonumber
\end{align}
where $T^{\pm}=T^{\pm}(x^{\pm})$. The $\cdots$ terms are of lowest order and
do not contribute to the surface integrals. Hence, the tentative boundary
conditions (\ref{Asympt relaxed metric mu}) are invariant under the full
conformal group in two dimensions, generated by $T^{+}(x^{+})$ and
$T^{-}(x^{-})$.

\section{Surface terms}

\setcounter{equation}{0}

\label{Surface Terms}

\subsection{Form of surface terms}

\label{Form of surface terms}

The conserved charges are computed within the canonical formalism,
\textquotedblleft\`{a} la Regge-Teitelboim\textquotedblright%
\ \cite{Regge-Teitelboim}. The Hamiltonian formalism for topologically massive
gravity is reviewed in appendix \ref{appendixA}.

The searched-for charges that generate the diffeomorphisms (\ref{Asympt KV})
must take the form \cite{Regge-Teitelboim}
\begin{equation}
H[\eta]=\hbox{``Bulk piece"}+Q_{+}[T^{+}]+Q_{-}[T^{-}]\,, \label{generator}%
\end{equation}
where the bulk piece is a linear combination of the constraints with
coefficients involving $\eta^{+},\eta^{-},\eta^{r}$ which has been explicitly
worked out in \cite{Carlip} and which are given in the appendices, and where
$Q_{+}[T^{+}]$ and $Q_{-}[T^{-}]$ are surface integrals at infinity that
involve only the asymptotic form of the vector field $\eta^{+},\eta^{-}%
,\eta^{r}$. On-shell, the bulk piece vanishes and $H[\eta]$ reduces to
$Q_{+}[T^{+}]+Q_{-}[T^{-}]$. As explained in \cite{Regge-Teitelboim}, the
variation of the surface integrals at infinity must cancel, under the given
boundary conditions, the surface terms that one picks up upon integrations by
parts in the bulk term. So, what is determined by the formalism are the
variations $\delta Q_{\pm}$ of the surface terms.

A crucial consistency requirement on the boundary conditions is that these
variations $\delta Q_{\pm}$ should be integrable, i.e., should define exact
forms in field space. If they are not, one must strengthen the boundary
conditions so as to fulfill this requirement.

The variation of the surface integrals giving the conserved charges under the
boundary conditions (\ref{Asympt relaxed metric mu}) are shown in appendix
\ref{appendixB} to be equal to
\begin{equation}
\delta Q_{\pm}=\left(  1\pm\frac{1}{\mu l}\right)  \delta Q_{\pm}^{0}+\delta
Q_{\pm}^{nl}\ , \label{deltaQQQQ}%
\end{equation}
where
\begin{equation}
\delta Q_{\pm}^{0}:=\frac{2}{l}\int T^{\pm}\delta f_{\pm\pm}d\phi\ ,
\label{deltaQ0}%
\end{equation}
is the standard contribution that one finds with the asymptotic behavior of
\cite{Brown:1986nw}, and
\begin{equation}
\delta Q_{\pm}^{nl}=\pm\frac{1}{2l}\left(  \mu^{2}l^{2}-1\right)  \int T^{\pm
}\left[  \left(  \frac{3}{\mu l}\mp1\right)  h_{\pm\pm}\delta h_{\mp\mp
}+\left(  \frac{1}{\mu l}\mp1\right)  h_{\mp\mp}\delta h_{\pm\pm}\right]
d\phi\ , \label{deltaQg}%
\end{equation}
is the additional nonlinear contribution coming from the relaxed terms
containing $h_{\mu\nu}$.

Now, while the standard contribution is integrable, the additional term is not
if we assume no restriction on the space of the $h_{\mu\nu}$'s. Indeed, one
gets that the second variation of the additional piece is given by
\[
\delta^{2}Q_{\pm}^{nl}\sim\delta h_{++}\wedge\delta h_{--}\ ,
\]
which is non-zero unless we assume a functional dependence $h_{++}%
=h_{++}(h_{--})$ (or $h_{--}=h_{--}(h_{++})$). As mentioned in the
introduction, the same situation is encountered when a scalar field is coupled
to gravity. There are two branches for the scalar field, behaving
asymptotically as
\[
\frac{a_{\pm}}{r^{\lambda_{\pm}}}\ .
\]
Integrability of the anti-de Sitter charges forces a functional relation
between $a_{+}$ and $a_{-}$.

\subsection{Constraints from asymptotic conformal invariance}

The functional relation must be chosen to be invariant under the asymptotic
symmetry. This is a non-trivial requirement because $h_{++}$ and $h_{--}$
transform differently.

Under the action of the Virasoro symmetry, one obtains%
\begin{align}
\delta_{\eta}h_{++}  &  =\frac{1}{2}\left[  \left(  3-\mu l\right)
\partial_{+}T^{+}-\left(  \mu l+1\right)  \partial_{-}T^{-}\right]
h_{++}+T^{-}\partial_{-}h_{++}+T^{+}\partial_{+}h_{++}\ ,\label{deltah++}\\
\delta_{\eta}h_{--}  &  =\frac{1}{2}\left[  (\mu l-1)\partial_{+}T^{+}+\left(
3+\mu l\right)  \partial_{-}T^{-}\right]  h_{--}+T^{-}\partial_{-}h_{--}%
+T^{+}\partial_{+}h_{--}\ . \label{deltah--}%
\end{align}
The searched-for relation $h_{++}=h_{++}(h_{--})$ must be consistent with
these equations, i.e., one must have%
\begin{equation}
\delta_{\eta}h_{++}=\frac{\delta h_{++}}{\delta h_{--}}\delta_{\eta}h_{--}\ .
\label{consistencyhs}%
\end{equation}
This has to be true for each Virasoro copy.

Considering first the right copy (generated by $T^{+}$), one gets
\begin{equation}
\left(  \left(  3-\mu l\right)  h_{++}+\left(  1-\mu l\right)  \frac{\delta
h_{++}}{\delta h_{--}}h_{--}\right)  \partial_{+}T^{+}=0\ , \label{condition1}%
\end{equation}
which integrates into
\begin{equation}
h_{++}=a_{0}^{+}\ (h_{--})^{\frac{3-\mu l}{\mu l-1}}\ , \label{Inv Right}%
\end{equation}
Similarly, one finds for the left copy (generated by $T^{-}$)
\begin{equation}
\left(  (1+\mu l)h_{++}+(3+\mu l)\frac{\delta h_{++}}{\delta h_{--}}%
h_{--}\right)  \partial_{-}T^{-}=0\ , \label{condition2}%
\end{equation}
which integrates into
\begin{equation}
h_{++}=a_{0}^{-}\ (h_{--})^{-\frac{1+\mu l}{\mu l+3}}\ . \label{Inv Left}%
\end{equation}
In these equations, $a_{0}^{\pm}$ are integration constants. As was to be
expected, conditions (\ref{Inv Right}) and (\ref{Inv Left}) are interchanged
under $\mu\longleftrightarrow-\mu$ and $x^{+}\longleftrightarrow x^{-}$.

\subsection{Conformal symmetry at infinity}

\label{Conformal symmetry}

As there is no value of $\mu$ for which the powers of $h_{--}$ in the
equations (\ref{Inv Right}) and (\ref{Inv Left}) are equal, the full conformal
invariance at infinity is absent, unless one sets the integration constants
equal to zero or infinity; i.e., only when one chirality is switched off. This
was the case considered in \cite{Henneaux:2009pw}. (Note that when $h_{\pm\pm
}$ is switched off, $h_{r_{\pm}}$ can be gauged away.) This is in sharp
contrast with what is found in the scalar field case where the functional
relation $a_{+}(a_{-})$ can be chosen to be invariant without removing one of
the two branches.

Now, when one of the chiralities is set to zero, the extra nonlinear
contribution (\ref{deltaQg}) vanishes and the charges reduce to the ones
obtained for the standard asymptotic conditions, namely,
\begin{equation}
Q_{\pm}[T^{\pm}]=\frac{2}{l}\left(  1\pm\frac{1}{\mu l}\right)  \int T^{\pm
}f_{\pm\pm}d\phi. \label{once}%
\end{equation}
Therefore, somewhat unexpectedly, the charges acquire no correction due to the
relaxed terms in the asymptotic expansion. Since these terms cannot be gauged
away, they can be viewed as \textquotedblleft massive graviton
hair\textquotedblright. Note in particular that the charges above for the
pp-waves (with $\phi$ identified) are all zero.

Once the generators of the asymptotic symmetries have been found, one can
compute their algebra applying the general theorems of \cite{Brown-Henneaux2}.
These guarantee that the algebra is the algebra of the conformal group with
possible central charges. The central charges are easily found from the
inhomogeneous transformation terms (under the conformal symmetry) in the
variations of the functions $f_{\pm\pm}$ that determine the charges. One
finds,
\begin{align}
\delta_{\eta}f_{++}  &  =\cdots-l^{2}\left(  \partial_{+}T^{+}+\partial
_{+}^{3}T^{+}\right)  \!/2\ ,\label{deltaf++}\\
\delta_{\eta}f_{--}  &  =\cdots-l^{2}\left(  \partial_{-}T^{-}+\partial
_{-}^{3}T^{-}\right)  \!/2\ , \label{deltaf--}%
\end{align}
where $\cdots$ stand for the homogeneous terms. One can thus infer that
$Q_{+}(T^{+})$ and $Q_{-}(T^{-})$ commute with each other and each fulfills
the Virasoro algebra with central charges
\begin{equation}
c_{\pm}=\left(  1\pm\frac{1}{\mu l}\right)  \,c\ , \label{c+-}%
\end{equation}
where $c$ is the central charge of \cite{Brown:1986nw},
\[
c=\frac{3l}{2G}\ .
\]

\subsection{Two chiralities present}

If one does not insist on conformal invariance at infinity, one can switch on
simultaneously both chiralities. If the relation between $h_{++}$ and $h_{--}$
is taken to be one of the two relations above, one preserves one chiral copy
of the Virasoro algebra. From the other chiral copy, only the zero mode
survives (the conditions (\ref{condition1}), (\ref{condition2}) which relate
$h_{++}$ to $h_{--}$ are trivially satisfied when $T^{\pm}$ are constants).
The asymptotic symmetry is then Virasoro$\times R$.

It is easy to integrate the charges. Let us assume first that we adopt the
relation (\ref{Inv Right}) between $h_{++}$ and $h_{--}$. Then, from Eq.
(\ref{deltaQQQQ}) one verifies that $\delta Q_{+}^{nl}=0$ and then
\begin{equation}
Q_{+}(T^{+})=\frac{2}{l}\left(  1+\frac{1}{\mu l}\right)  \int T^{+}%
f_{++}d\phi\ , \label{Q+conditionRight}%
\end{equation}
for the Virasoro algebra with central charge%
\begin{equation}
c_{+}=\left(  1+\frac{1}{\mu l}\right)  \,c\ ,
\end{equation}
and%
\begin{equation}
Q_{-}[T^{-}]=\frac{2}{l}\left(  1-\frac{1}{\mu l}\right)  \int T^{-}\left[
f_{--}-(1+\mu l)a_{0}^{+}\ h_{--}{}^{\frac{2}{\mu l-1}}\right]  d\phi\ ,
\label{Q-conditionRight}%
\end{equation}
for the generator of the zero mode ($\partial_{-}T^{-}=0$) (See appendices
\ref{appendixB} and \ref{AppendixC}).

Analogously, if one considers the relation (\ref{Inv Left}) between $h_{++}$
and $h_{--}$, one obtains%
\begin{equation}
Q_{-}(T^{-})=\frac{2}{l}\left(  1-\frac{1}{\mu l}\right)  \int T^{-}%
f_{--}d\phi\ , \label{Q-conditionLeft}%
\end{equation}
for the Virasoro algebra with central charge
\begin{equation}
c_{-}=\left(  1-\frac{1}{\mu l}\right)  \,c\ ,
\end{equation}
and%
\begin{equation}
Q_{+}(T^{+})=\frac{2}{l}\left(  1+\frac{1}{\mu l}\right)  \int T^{+}\left[
f_{++}-(1-\mu l)a_{0}^{-}\ (h_{--})^{\frac{2}{\mu l+3}}\right]  d\phi\ ,
\label{Q+conditionLeft}%
\end{equation}
for the generator of the zero mode ($\partial_{+}T^{+}=0$). Naturally, the
expressions for the charges and the central extension associated to
(\ref{Inv Left}) coincide with the ones for (\ref{Inv Right}) making
$\mu\longleftrightarrow-\mu$ and $x^{+}\longleftrightarrow x^{-}$.

One can also break further the Virasoro symmetry through arbitrary boundary
conditions. For a generic relation between $h_{++}$ and $h_{--}$ of the form%
\begin{equation}
h_{++}=w^{\prime}(h_{--})\ , \label{w'}%
\end{equation}
the conditions (\ref{condition1}), (\ref{condition2}) are fulfilled only for
the zero modes of both copies of the Virasoro symmetry (i.e., when $T^{+}$ and
$T^{+}$ are constants). Hence, the asymptotic symmetry is broken down to $R$
$\times U(1)$, and the charges acquire the form%
\begin{align}
Q_{+}[T^{+}]  &  =\frac{2}{l}\left(  1+\frac{1}{\mu l}\right)  \int
T^{+}\left[  f_{++}+\frac{1}{4}(\mu l-1)\left(  2w-(\mu l-1)w^{\prime}%
h_{--}\right)  \right]  d\phi\ ,\label{Q+Arbitrary}\\
Q_{-}[T^{-}]  &  =\frac{2}{l}\left(  1-\frac{1}{\mu l}\right)  \int
T^{-}\left[  f_{--}+\frac{1}{4}(\mu l+1)\left(  2w-(\mu l+3)w^{\prime}%
h_{--}\right)  \right]  d\phi\ , \label{Q-Arbitrary}%
\end{align}
with no possible central extension.

What we found in this section is somewhat reminiscent of what occurs for a
scalar field in $d$ dimensions with an arbitrary functional relation
$a_{+}(a_{-})$. Although the metric still has the same asymptotic AdS
invariance, the scalar field breaks then the symmetry down to $R\times
SO(d-1)$ because the relationship is not maintained under the action of an
asymptotic radial diffeomorphism \cite{Henneaux:2006hk}. This kind of breaking
of asymptotic AdS invariance for scalar fields has been considered in
\cite{Designer}, following ideas from the AdS/CFT correspondence \cite{Witten}.

\section{Conclusions}

\setcounter{equation}{0} In this paper we have presented an extended set of
boundary conditions that includes the ones recently proposed in
\cite{Henneaux:2009pw} with a slower decay at infinity than the one for pure
standard gravity discussed in \cite{Brown:1986nw}. This set considers
simultaneously both chiralities (below the chiral point), and it was shown
that requiring well-defined charges, whose variations are not only finite but
also integrable, forces one to impose a functional relationship between the
two chiralities. It was proved that there is no possible relation that
preserves the conformal symmetry at infinity, except setting one of the two
chiralities to zero; otherwise, the best that one can achieve is
Virasoro$\times R$. Therefore, the boundary conditions given in the earlier
treatment appear to be the best possible ones compatible with conformal symmetry.

It would be interesting to explore whether a similar treatment of the
asymptotic behavior for topologically massive gravity along the lines
presented here could be performed around warped AdS \cite{Anninos-Strominger},
\cite{Dio-Monica-Warped-AdS3}, \cite{Compere-Detournay}.

\acknowledgments

We thank D. Anninos, M. Becker, G. Giribet, D. Grumiller and N. Johansson for
useful discussions and comments. This research is partially funded by FONDECYT
grants N%
${{}^o}$
1085322, 1095098, 1100755, and by the Conicyt grant \textquotedblleft Southern
Theoretical Physics Laboratory\textquotedblright\ ACT-91. The work of MH is
partially supported by IISN - Belgium (conventions 4.4511.06 and 4.4514.08)
and by the Belgian Federal Science Policy Office through the Interuniversity
Attraction Pole P6/11. C. M. and R. T. wish to thank the kind hospitality at
the Physique th\'{e}orique et math\'{e}matique at the Universit\'{e} Libre de
Bruxelles and the International Solvay Institutes. The Centro de Estudios
Cient\'{\i}ficos (CECS) is funded by the Chilean Government through the
Millennium Science Initiative and the Centers of Excellence Base Financing
Program of Conicyt. CECS is also supported by a group of private companies
which at present includes Antofagasta Minerals, Arauco, Empresas CMPC, Indura,
Naviera Ultragas and Telef\'{o}nica del Sur. CIN is funded by Conicyt and the
Gobierno Regional de Los R\'{\i}os.

\appendix

\section{Hamiltonian Formulation}

\label{appendixA} \setcounter{equation}{0}

\subsection{Action in first order form}

The canonical analysis of topologically massive gravity has been performed in
\cite{Deser-Xiang, Carlip}. We shall follow very closely the procedure devised
in \cite{Carlip} to reach the Hamiltonian form of the theory. This procedure
has the advantage of bypassing many of the technical intricacies associated
with a more conventional application of the canonical formalism for theories
with higher order derivatives.

By introducing Lagrange multipliers for the torsion constraints and using as
new connection \cite{Giacomini-Troncoso-Willison}%
\begin{equation}
A^{a}=\omega^{a}+\mu\,e^{a}\,, \label{A}%
\end{equation}
one can rewrite the action in first order form as follows,
\begin{equation}
I[e,A,\beta]=\frac{1}{\mu}I_{CS}[A]+\int\left[  \beta^{a}\left(  De_{a}%
-\mu\epsilon_{abc}e^{b}\wedge e^{c}\right)  -\alpha\epsilon_{abc}e^{a}\wedge
e^{b}\wedge e^{c}\right]  \label{FOAction}%
\end{equation}
where $I_{CS}[A]$ reads
\begin{equation}
I_{CS}[A]=\int\left[  A^{a}\wedge\left(  dA_{a}+\frac{1}{3}\epsilon
_{abc}\,A^{b}\wedge A^{c}\right)  \right]
\end{equation}
and where the covariant derivative is taken with respect to the connection
$A$. The parameter $\alpha$ is given by
\begin{equation}
\alpha=\frac{1}{3}\left(  \mu^{2}-\frac{1}{l^{2}}\right)  .
\end{equation}
The independent variables are the components of the triad $e^{a}=e_{\;\mu}%
^{a}dx^{\mu}$, the components of the connection $A^{a}=A_{\;\mu}^{a}dx^{\mu}$
and the components of the Lagrange multiplier $\beta^{a}=\beta_{\;\mu}%
^{a}dx^{\mu}$.

The action (\ref{FOAction}) is already in the Hamiltonian form
\textquotedblleft$\int dt(p\dot{q}-\lambda^{\alpha}H_{\alpha}(q,p))$" where
the $p$'s and the $q$'s are the spatial components of the dynamical variables
and where the Lagrange multipliers are their temporal components. Explicitly,
one can write (\ref{FOAction}) as
\begin{align}
&  I[e_{\;i}^{a},A_{\;i}^{a},\beta_{\;i}^{a};e_{\;0}^{a},A_{\;0}^{a}%
,\beta_{\;0}^{a}]=\nonumber\\
&  \;\;\;\;\;\;\;\;\;\;\;\;\;\;\int d^{3}x\left[  \epsilon^{ij}\left(
-\frac{1}{\mu}A_{\;i}^{a}\,\dot{A}_{aj}-\beta_{\;i}^{a}\,\dot{e}_{aj}\right)
-A_{\;0}^{a}J_{a}-\beta_{\;0}^{a}T_{a}-e_{\;0}^{a}B_{a}\right]
\label{FOHAction}%
\end{align}
with
\begin{align}
J_{a}  &  =-\frac{2}{\mu}\epsilon^{ij}\left(  F_{aij}+\frac{\mu}{2}%
\epsilon_{abc}\,\beta_{\;i}^{b}\,e_{\;j}^{c}\right) \\
T_{a}  &  =-\epsilon^{ij}\left(  D_{i}e_{aj}-\mu\,\epsilon_{abc}\,e_{\;i}%
^{b}\,e_{\;j}^{c}\right) \\
B_{a}  &  =-\epsilon^{ij}\left(  D_{i}\beta_{aj}-2\,\mu\,\epsilon_{abc}%
\,\beta_{\;i}^{b}\,e_{\;j}^{c}-3\,\alpha\,\epsilon_{abc}\,e_{\;i}^{b}%
\,e_{\;j}^{c}\right)
\end{align}
and $F_{a}=dA_{a}+\frac{1}{2}\epsilon_{abc}\,A^{b}\wedge A^{c}$.

The kinetic term in the action (\ref{FOHAction}) implies the following Poisson
brackets among the variables
\begin{align}
&  [A^{a}_{\; i}, A^{b}_{\; j}] = \frac{\mu}{2} \eta^{ab} \epsilon_{ij}\\
&  [e^{a}_{\; i}, \beta^{b}_{\; j}] = \eta^{ab} \epsilon_{ij}%
\end{align}

\subsection{Constraints}

Varying the Lagrange multipliers yield the constraints $J_{a}\approx0$,
$T_{a}\approx0$ and $B_{a}\approx0$. Their Poisson brackets have been computed
in \cite{Carlip}. For completeness, we reproduce them here,%
\begin{align}
\left[  J[\xi],C[\eta]\right]   &  =-C[\xi\times\eta]\label{bJJ}\\
\left[  T[\xi],T[\eta]\right]   &  =-\frac{\mu}{2}\int d^{2}x\,\xi^{a}\eta
^{b}\left(  \epsilon^{ij}e_{ai}e_{bj}\right) \label{bTT}\\
\left[  B[\xi],T[\eta]\right]   &  =-\frac{\mu}{2}J[\xi\times\eta]+2\mu
T[\xi\times\eta]+\frac{\mu}{2}\int d^{2}x\,\xi^{a}\eta^{b}\left(
\epsilon^{ij}\beta_{ai}e_{bj}-\eta_{ab}\epsilon^{ij}\beta_{\;i}^{c}%
e_{cj}\right) \label{bBT}\\
\left[  B[\xi],B[\eta]\right]   &  =2\mu B[\xi\times\eta]+6\alpha T[\xi
\times\eta]-\frac{\mu}{2}\int d^{2}x\,\xi^{a}\eta^{b}\left(  \epsilon
^{ij}\beta_{ai}\beta_{bj}\right)  \label{bBB}%
\end{align}
Here, the notation $C[\xi]$ stands for $\int d^{2}x\,\xi^{a}C_{a}$ for any
constraint $C_{a}\equiv J_{a}$, $T_{a}$ or $B_{a}$, where $\xi^{a}$ are
arbitrary parameters assumed for the moment to have compact support in order
to avoid surface terms at infinity (dealt with below). The notation $\xi
\times\eta$ is short for $(\xi\times\eta)^{a}=\epsilon^{abc}\xi_{b}\eta_{c}$.
The first bracket (\ref{bJJ}) simply follows from the fact that the constraint
$J[\xi]$ is the generator of local Lorentz transformations.

The brackets between the constraints imply in particular the following
equation for $\dot{T_{a}}$
\begin{equation}
\dot{T_{a}} \approx- \frac{\mu}{2} \beta^{b}_{\; 0} \epsilon^{ij} e_{ai}
e_{bj}- \frac{\mu}{2} e^{b}_{\; 0} \epsilon^{ij} \beta_{bi} e_{aj} + \frac
{\mu}{2} e_{a0} \Delta\label{ConsistencyForT}%
\end{equation}
where
\begin{equation}
\Delta= \epsilon^{ij} \, \beta^{c}_{\; i} \, e_{cj}.
\end{equation}

To analyse the nature of the constraints and determine whether they imply
further constraints, it is convenient to introduce the components $n_{a}$ in
the triad frame of the normal to the hypersurfaces $x^{0} =$ const. These are
defined through
\begin{equation}
n^{a} e_{ai} = 0, \; \; \; \; n^{a} n_{a} = -1
\end{equation}
($n^{a} e_{ai} \equiv n_{i} = 0$). As observed in \cite{NeTe}, the $n^{a}$'s
are functions of the canonical variables $e_{ai}$ only and do not depend on
the Lagrange multipliers. Explicitly, $n^{a} = u^{a} / \sqrt{-u^{b} u_{b}}$
with $u^{a} = -\epsilon^{abc} \epsilon^{ij} e_{ai} e_{bj}$. Note that $n^{a}
e_{a0} = n_{0} \not = 0$. Some useful relations are
\begin{align}
&  e_{ai} e^{a}_{\; j} = g_{ij}\\
&  e_{b}^{\;i} e_{ai} = \eta_{ab} + n_{a} n_{b}\\
&  \epsilon^{ij} e_{ai} = \sqrt{g} \epsilon_{abc} e^{bj} n^{c}%
\end{align}
where $e_{b}^{\;i} = g^{ij} e_{bj}$ with $g^{ij}$ the inverse of the spatial
two-dimensional metric $g_{ij}$ and $g = \det(g_{ij})$. From the last
relation, one derives
\begin{equation}
\epsilon^{ij} e_{ai} e_{bj} = \sqrt{g} \epsilon_{abc} n^{c}%
\end{equation}
and
\begin{equation}
\epsilon^{abc} \epsilon^{ij} e_{ai} e_{bj} n_{c} = 2 \sqrt{g}%
\end{equation}

Projecting (\ref{ConsistencyForT}) along the normal yields $n^{a} \dot{T_{a}}
\approx n^{a} e_{a0} \Delta$. Hence, since the constraints must be preserved
in time and since $n^{a} e_{a0} \not = 0$, one gets the further constraint
\begin{equation}
\Delta\approx0.
\end{equation}

The full set of constraints at this stage is given by $\{J_{a},T_{a}%
,B_{a},\Delta\}\approx0$. To verify that there is no other constraint and to
separate the constraints into first class and second class, it is convenient
to redefine the constraint $B_{a}$ as \cite{Carlip}
\begin{equation}
\hat{B}[\xi]=B[\xi]+T[\hat{\xi}]\ ,
\end{equation}
where
\begin{equation}
\hat{\xi}^{a}=e^{aj}\beta_{bj}\xi^{b}+fn^{a}\ ,
\end{equation}
and $f$ is given by
\begin{equation}
f=-\beta_{ai}n^{a}e^{ci}\xi_{c}+\left(  \beta_{ai}e^{ai}+\frac{9\alpha}{\mu
}\right)  n^{c}\xi_{c}\ .
\end{equation}
One easily verifies that $e_{ai}\hat{\xi}^{a}=\beta_{ai}\xi^{a}$ and that the
constraints $\hat{B}_{a}\approx0$ are first class. Our expression for
$\hat{\xi}^{a}$ can be checked to coincide with the one given in
\cite{Carlip}, but we have rewritten it in a way that makes it clear that the
redefinition of the constraints $B_{a}\rightarrow\hat{B}_{a}$ involves only
the canonical variables \emph{and not the Lagrange multipliers $e_{\;t}^{a}$
or $\beta_{\;t}^{a}$ (which would not be permissible)}.

One can in fact verify that
\begin{equation}
h[\xi]=\int d^{2}x\left(  \xi^{a}\hat{B}_{a}+A_{\;\mu}^{a}\xi^{\mu}%
J_{a}\right)
\end{equation}
generates on shell the Lie derivatives of the canonical variables, i.e.,
\begin{equation}
\left[  X,h[\xi]\right]  =-L_{\xi}X
\end{equation}
modulo terms that vanish when the equations of motion hold \cite{Carlip}.
Here, $\xi^{a}$ is related to the vector field $\xi^{\mu}$ parametrizing the
infinitesimal diffeomorphism through $\xi^{a}=e_{\;\mu}^{a}\xi^{\mu}$.

While the constraints $\{J_{a},\hat{B}_{a}\}$ are first class, the remaining
constraints $\{T_{a},\Delta\}$ are second class. Their (invertible) brackets
are explicitly computed in \cite{Carlip}, to which we refer. There are thus 6
first class constraints and 4 second class constraints. Given that the number
of canonical variables is 18, this gives $18-2\times6-4=2$ physical canonical
variables, corresponding to one conjugate canonical pair -- and hence one
physical degree of freedom -- (per space point), independently of the value of
$\mu$.

\section{Surface integrals}

\label{appendixB} \setcounter{equation}{0}

We now turn to the discussion of the surface terms that must be added to the
diffeomorphism generators under the boundary conditions given in the text.

These boundary conditions have been written in terms of the metric and our
first task is to rewrite them in terms of the triads since the canonical
formalism has been developed in terms of these. To that end, we find it
convenient to freeze the local Lorentz gauge freedom at infinity by imposing
that the triads are asymptotically given by
\begin{equation}
e_{\;\mu}^{a}=\bar{e}_{\;\mu}^{a}+\Delta e_{\;\mu}^{a} \label{e}%
\end{equation}
where the $\bar{e}_{\;\mu}^{a}$'s are the following choice of AdS triads%
\begin{align}
\bar{e}^{0}  &  =\left(  1+\frac{r^{2}}{l^{2}}\right)  ^{1/2}\ dt=\frac{l}%
{2}\left(  1+\frac{r^{2}}{l^{2}}\right)  ^{1/2}\ \left(  dx^{+}+dx^{-}\right)
\ ,\nonumber\\
\bar{e}^{1}  &  =\left(  1+\frac{r^{2}}{l^{2}}\right)  ^{-1/2}%
\ dr\ ,\label{e bar}\\
\bar{e}^{2}  &  =r\ d\phi=\frac{1}{2}r\ \left(  dx^{+}-dx^{-}\right)
\ ,\nonumber
\end{align}
and the perturbation $\Delta e_{\;\mu}^{a}$ is related to the metric
perturbation $\Delta g_{\mu\nu}$ through
\begin{equation}
\Delta e_{\;\mu}^{a}=\bar{e}_{\;\rho}^{a}\bar{g}^{\rho\sigma}\left(  \frac
{1}{2}\Delta g_{\sigma\mu}-\frac{1}{8}\Delta g_{\sigma\gamma}\bar{g}%
^{\gamma\beta}\Delta g_{\beta\mu}\right)  \ . \label{deltae}%
\end{equation}
Note that in the case of asymptotic conditions invariant under the full
conformal symmetry, for which one chirality is switched off, only the linear
term in the metric perturbation is actually required.

The ordinary Lie derivative of the triads does not preserve this gauge-fixing
of the triads. More precisely, when performing a diffeomorphism generated by a
vector field $\xi^{\mu}$ that approaches at infinity an asymptotic symmetry
according to (\ref{Asympt KV}), one must compensate by a local Lorentz
transformation that brings one back to that chosen gauge. Explicitly, the
parameter of that compensating Lorentz transformation is found to be
\begin{equation}
-A^{a}_{\mu}\xi^{\mu}+ \left(  \mu\pm\frac{1}{\ell}\right) e^{a}_{\mu}\xi
^{\mu}_{(\pm)}%
\end{equation}
where $\xi_{(\pm)}^{\mu}$ are the pieces of $\xi^{\mu}$ asymptotically
determined by $T^{+}$ and $T^{-}$, respectively. Hence, the full generators,
containing the diffeomorphism part plus the compensating Lorentz
transformation, read
\begin{equation}
H[\xi_{(\pm)}]=\hat{B}[\xi_{(\pm)}]+\left(  \mu\pm\frac{1}{\ell}\right)
J[\xi_{(\pm)}]+\hbox{ surface terms}
\end{equation}
where $\xi_{(\pm)}^{a} = e^{a}_{\mu}\xi_{(\pm)}^{\mu}$.

Up to the surface terms, $H[\xi_{(\pm)}]$ coincides with the $L_{\pm}[\xi]$ of
reference \cite{Carlip}, but we have followed a different logic to arrive at
that expression. We can thus take over from that reference the computation of
the terms that arise upon integration by parts in the bulk piece of
$H[\xi_{(\pm)}]$. One finds
\begin{equation}
\delta Q_{\pm}=\int_{\partial\Sigma}\xi_{(\pm)}^{\mu}\left[  e_{\;\mu}%
^{a}\delta\beta_{a\phi}+\beta_{\;\mu}^{a}\delta e_{a\phi}+2\left(  1\pm
\frac{1}{\mu\ell}\right)  e_{\;\mu}^{a}\delta A_{a\phi}\right]  d\phi\ .
\label{surfaceappendix}%
\end{equation}
After straightforward algebra, one gets from (\ref{surfaceappendix}) the
expression (\ref{deltaQQQQ}) for $\delta Q_{\pm}$.

This can be seen as follows: The field equation that comes from the variation
with respect to $\beta_{\ \mu}^{a}$ implies the vanishing of the torsion,
which allows to express the spin connection $\omega_{\ \mu}^{a}$ in terms of
the triad (given by Eq. (\ref{e}), with (\ref{e bar}) and (\ref{deltae})),
hence determining the field $A_{\ \mu}^{a}$ defined in (\ref{A}). Analogously,
varying with respect to $A_{\ \mu}^{a}$ gives an algebraic equation that
allows to find $\beta_{\ \mu}^{a}$, which is given by%
\[
\beta_{\ \mu}^{a}=-\frac{2}{\mu}e_{\ \nu}^{a}\left(  R_{\ \mu}^{\nu}-\frac
{1}{4}\delta_{\ \mu}^{\nu}R+\frac{\mu^{2}}{2}\delta_{\ \mu}^{\nu}\right)  \ .
\]
The asymptotic form of the field equation $E_{rr}=0$ in (\ref{Err}), which
turns out to be a constraint, is also useful in order to express the terms
appearing $\delta Q_{\pm}$, given by%
\begin{align*}
\xi_{\pm}^{\mu}e_{\;\mu}^{a}\delta\beta_{a\phi}  &  =\pm\frac{1}{4\mu l^{2}%
}\left(  \mu^{2}l^{2}-1\right)  T^{\pm}\left[  2\delta h_{\pm\pm}\ r^{1\pm\mu
l}+2\delta f_{+-}-2\delta f_{\pm\pm}+11h_{\pm\pm}\delta h_{\mp\mp}+9h_{\mp\mp
}\delta h_{\pm\pm}\right] \\
& \\
\xi_{\pm}^{\mu}\beta_{\;\mu}^{a}\delta e_{a\phi}  &  =\pm\frac{1}{4\mu l^{2}%
}\left(  \mu^{2}l^{2}-1\right)  T^{\pm}\left[  -2\delta h_{\pm\pm}\ r^{1\pm\mu
l}+2\delta f_{+-}-2\delta f_{\pm\pm}+3h_{\pm\pm}\delta h_{\mp\mp}+h_{\mp\mp
}\delta h_{\pm\pm}\right] \\
& \\
\xi_{\pm}^{\mu}e_{\;\mu}^{a}\delta A_{a\phi}  &  =\frac{1}{4l}T^{\pm}\left[
2(1\pm\mu l)\delta f_{\pm\pm}+(1\mp\mu l)\left[  2\delta f_{+-}+\left(
4\pm\mu l\right)  \delta(h_{++}h_{--})\right]  \right]
\end{align*}
Therefore, the divergences coming from the first and second terms above cancel
out, and the variation of the charge in (\ref{surfaceappendix}) reduces to%
\[
\delta Q_{\pm}=\left(  1\pm\frac{1}{\mu l}\right)  \delta Q_{\pm}^{0}+\delta
Q_{\pm}^{nl}\ ,
\]
whre $\delta Q_{\pm}^{0}$ and $\delta Q_{\pm}^{nl}$, are given in Eqs.
(\ref{deltaQ0}) and (\ref{deltaQg}), respectively.

\section{Asymptotic form of the field equations}

Once the auxiliary fields $A_{\ \mu}^{a}$ and $\beta_{\ \mu}^{a}$ are
eliminated, the field equations coming from (\ref{actionTMG}) read%
\begin{equation}
E_{\mu\nu}:=G^{\mu}{}_{\sigma}-\frac{1}{l^{2}}\delta_{\sigma}^{\mu}-\frac
{1}{\mu}C^{\mu}{}_{\sigma}=0\ ,\label{eom}%
\end{equation}
where $\mu\neq0$ is the mass parameter, $l$ is the AdS radius, and $C^{\mu}%
{}_{\sigma}:=\left(  -g\right)  ^{-1/2}\epsilon^{\mu\nu\rho}\nabla_{\nu
}\left(  R_{\rho\sigma}-\frac{1}{4}g_{\rho\sigma}R\right)  $, stands for the
Cotton tensor.

In the case of $|\mu l|<1$, for the asymptotic conditions given by
(\ref{Asympt relaxed metric mu}), the leading terms of the relevant field
equations reduce to%
\begin{align}
E_{rr}  &  =r^{-4}\left[  -l^{-2}f_{rr}+4f_{+-}+(5-\mu^{2}l^{2})h_{++}%
h_{--}\right]  +\cdot\cdot\cdot\label{Err}\\
E_{r\pm}  &  =\frac{1}{2}\left(  1\pm\frac{1}{\mu l}\right)  r^{-3}\left[
4\partial_{\mp}f_{\pm\pm}-\left(  1\mp\mu l\right)  \left[  (1\mp\mu
l)h_{\mp\mp}\partial_{\pm}h_{\pm\pm}+\left(  3\mp\mu l\right)  h_{\pm\pm
}\partial_{\pm}h_{\mp\mp}\right]  \right]  +\cdot\cdot\cdot\label{Er+-}%
\end{align}
where the first equation was used to simplify the second one. The remaining
field equations are of subleading orders as compared with the asymptotic
behaviour of the metric.

Note that, by integrating the equations from infinity, there is no apparent
obstruction coming from (\ref{Err}) and (\ref{Er+-}) in order to construct
solutions where both chiralities are present. Indeed, one could choose
$h_{++}$ and $h_{--}$ to be independent functions of $x^{+}$\ and $x^{-}$;
however, as explained in Section \ref{Form of surface terms}, integrability of
the surface generators requires them to be functionally dependent. For a
generic choice of $h_{++}(h_{--})$, the asymptotic symmetry is broken down to
$R\times U(1)$, while if one of the relations (\ref{Inv Right}) or
(\ref{Inv Left}) are fulfilled, then the symmetry enhances to Virasoro$\times
R$. Full conformal symmetry at infinity is recovered when a single chirality
is switched on. In this case, there is no obstruction for solutions
generalizing the pp-waves to exist ($h_{++}$ or $h_{--}$ depend explicitly on
$x^{+}$ and $x^{-}$ ). And indeed, this is realized for the class of metrics
recently found in \cite{Chow:2009km}, \cite{Kundt-Spacetimes},
\cite{Chow:2009vt}.

\section{Asymptotic symmetries and canonical generators algebra}

\label{AppendixC}

\subsection{Full conformal symmetry}

When one chirality is switched off (i.e., when $h_{++}$ or $h_{--}$ vanishes),
as explained in Section \ref{Conformal symmetry}, the asymptotic symmetries
correspond to the conformal group in two dimensions generated by
(\ref{Asympt KV}). Then, $f_{++}$ and $f_{--}$ are straightforwardly found to
transform as
\begin{align}
\delta_{\eta}f_{++}  &  =2f_{++}\partial_{+}T^{+}+T^{-}\partial_{-}%
f_{++}+T^{+}\partial_{+}f_{++}-l^{2}\left(  \partial_{+}T^{+}+\partial_{+}%
^{3}T^{+}\right)  \!/2\ ,\label{delta f++ con tutti}\\
\delta_{\eta}f_{--}  &  =2f_{--}\partial_{-}T^{-}+T^{-}\partial_{-}%
f_{--}+T^{+}\partial_{+}f_{--}-l^{2}\left(  \partial_{-}T^{-}+\partial_{-}%
^{3}T^{-}\right)  \!/2\ . \label{delta f-- con tutti}%
\end{align}
From Eq. (\ref{Er+-}), one verifies that, on shell%
\begin{equation}
\partial_{+}f_{--}=0=\partial_{-}f_{++}\ ,
\end{equation}
and so (\ref{delta f++ con tutti}) and (\ref{delta f-- con tutti}) reduce to
\begin{align}
\delta_{\eta}f_{++}  &  =2f_{++}\partial_{+}T^{+}+T^{+}\partial_{+}%
f_{++}-l^{2}\left(  \partial_{+}T^{+}+\partial_{+}^{3}T^{+}\right)  \!/2\ ,\\
\delta_{\eta}f_{--}  &  =2f_{--}\partial_{-}T^{-}+T^{-}\partial_{-}%
f_{--}-l^{2}\left(  \partial_{-}T^{-}+\partial_{-}^{3}T^{-}\right)  \!/2\ .
\end{align}
Therefore, the variation of the charges (\ref{once}) reads%
\begin{align*}
\delta_{\eta_{2}}Q_{\pm}[T_{1}^{\pm}]  &  =[Q_{\pm}[T_{1}^{\pm}],Q_{+}%
[T_{2}^{+}]+Q_{-}[T_{2}^{-}]]\\
&  =\frac{2}{l}\left(  1\pm\frac{1}{\mu l}\right)  \int T_{1}^{\pm}%
\ \delta_{\eta_{2}}f_{\pm\pm}d\phi\ ,\\
&  =\frac{2}{l}\left(  1\pm\frac{1}{\mu l}\right)  \int d\phi\left[
(T_{1}^{\pm}\partial_{\pm}T_{2}^{\pm}-T_{2}^{\pm}\partial_{\pm}T_{1}^{\pm
})f_{\pm\pm}-\frac{l^{2}}{2}T_{1}^{\pm}\left(  \partial_{\pm}T_{2}^{\pm
}+\partial_{\pm}^{3}T_{2}^{\pm}\right)  \right] \\
&  =Q_{\pm}\left[  [T_{1}^{\pm},T_{2}^{\pm}]\right]  -\left(  1\pm\frac{1}{\mu
l}\right)  \ l\int d\phi\ T_{1}^{+}\left(  \partial_{+}T_{2}^{+}+\partial
_{+}^{3}T_{2}^{+}\right)  \ ,
\end{align*}
which implies that $Q_{+}$ and $Q_{-}$ commute with each other and each
fulfills the Virasoro agebra with central charges
\begin{equation}
c_{\pm}=\left(  1\pm\frac{1}{\mu l}\right)  \,c
\end{equation}
(see \cite{Brown-Henneaux2} for general theorems).

\subsection{Two chiralities present}

In the case of $|\mu l|<1$, one obtains quite generally that $f_{++}$ and
$f_{--}$ respectively transform as in Eqs. (\ref{delta f++ con tutti}) and
(\ref{delta f-- con tutti}) under the action of the Virasoro algebra, and
\begin{align*}
\delta_{\eta}h_{++}  &  =\frac{1}{2}\left[  \left(  3-\mu l\right)
\partial_{+}T^{+}-\left(  \mu l+1\right)  \partial_{-}T^{-}\right]
h_{++}+T^{-}\partial_{-}h_{++}+T^{+}\partial_{+}h_{++}\ ,\\
\delta_{\eta}h_{--}  &  =\frac{1}{2}\left[  (\mu l-1)\partial_{+}T^{+}+\left(
3+\mu l\right)  \partial_{-}T^{-}\right]  h_{--}+T^{-}\partial_{-}h_{--}%
+T^{+}\partial_{+}h_{--}\ .
\end{align*}

As we have seen, when both chiralities are present, only one copy of the
Virasoro algebra actually survives as asymptotic symmetry. Let us assume first
that we adopt the relation (\ref{Inv Right}) between $h_{++}$ and $h_{--}$.
Thus, from Eq. (\ref{Q+conditionRight}) one obtains%
\[
\delta_{\xi_{2}}Q_{+}[\xi_{1}]=\frac{2}{l}\left(  1+\frac{1}{\mu l}\right)
\int T_{1}^{+}\delta_{\xi_{2}}f_{++}d\phi\ .
\]
Remarkably, by virtue of (\ref{Inv Right}), the nonlinear terms of the field
Eq. $E_{r+}=0$ in (\ref{Er+-}) vanish, so that it reduces to%
\[
\partial_{-}f_{++}=0\ ,
\]
and hence $Q_{+}$ fulfills the Virasoro algebra with the corresponding central
charge $c_{+}$.

In the case of $Q_{-}$, from (\ref{Q-conditionRight}) one obtains%
\[
\delta_{\xi_{2}}Q_{-}[\xi_{1}]=\frac{2}{l}\left(  1-\frac{1}{\mu l}\right)
\int T_{1}^{-}\left[  \delta_{\xi_{2}}f_{--}-2a_{0}^{+}\ \frac{\mu l+1}{\mu
l-1}(h_{--})^{\frac{3-\mu l}{\mu l-1}}\delta_{\xi_{2}}h_{--}\right]  d\phi\ ,
\]
where $T_{1}^{-}$ and $T_{2}^{-}$ are constants. Since $\delta_{\xi_{2}}%
f_{--}$ and $\delta_{\xi_{2}}h_{--}$ can be read from
(\ref{delta f++ con tutti}) and (\ref{deltah--}), respectively, the variation
of the charge reads%
\begin{equation}
\delta_{\xi_{2}}Q_{-}[\xi_{1}]=\frac{2}{l}\left(  1+\frac{1}{\mu l}\right)
\int d\phi\ T_{1}^{-}T_{2}^{+}\ \partial_{+}f_{--}\ . \label{deltaQ-}%
\end{equation}
In this case, once one uses the relation (\ref{Inv Right}), the nonlinear
terms of the field Eq. $E_{r-}=0$ in (\ref{Er+-}) do not vanish, so that it
reads%
\[
\partial_{+}f_{--}=2a_{0}^{+}\ \frac{\mu l+1}{\mu l-1}(h_{--})^{\frac{3-\mu
l}{\mu l-1}}\partial_{-}h_{--}\ .
\]
Thus, plugging this field equations into (\ref{deltaQ-}) and integrating by
parts, one obtains that%
\[
\delta_{\xi_{2}}Q_{-}[\xi_{1}]=0\ .
\]

It is simple to verify that for the other condition in Eq. (\ref{Inv Left}),
the corresponding charge (\ref{Q-conditionLeft}) generates a Virasoro algrabra
with central extension $c_{-}$ and commutes with the abelian one in
(\ref{Q+conditionLeft}). Indeed, this corresponds to making $\mu
\longleftrightarrow-\mu$ and $x^{+}\longleftrightarrow x^{-}$ in the
computations of this section.

\bigskip

Analogously, it is very simple to treat the case of an arbitrary relation
between $h_{++}$ and $h_{--}$. Since the conditions (\ref{condition1}),
(\ref{condition2}) are fulfilled only for the zero modes of both copies of the
Virasoro symmetry, the algebra of the canonical generators coincides then with
the one of the remaining asymptotic symmetries, $R$ $\times U(1)$, with no
central extension.

\section{Chiral point}

For completeness, we derive below the formulas relevant to the chiral point
$|\mu l|=1$. This is somewhat out of the main line of our paper which explores
the possibility (only available for $|\mu l|<1$) of switching on
simultaneously both chiralities. But since these formulas, announced in
\cite{Henneaux:2009pw}, are easily derived from the above computations, we
provide them here. We assume $\mu l=1$. The case $\mu l=-1$ just corresponds
to the interchange $x^{+}\longleftrightarrow x^{-}$.

\subsection{Surface integrals}

In the case $\mu l=1$, only the negative chirality is present (i.e.,
$h_{++}=h_{r+}=0$). The suitable asymptotic behaviour for $\Delta g_{\mu\nu}$
possessing full conformal invariance at infinity is given by
\cite{Henneaux:2009pw}%
\begin{equation}%
\begin{array}
[c]{lll}%
\Delta g_{rr} & = & f_{rr}r^{-4}+\cdot\cdot\cdot\\
\Delta g_{r+} & = & f_{r+}r^{-3}+\cdot\cdot\cdot\\
\Delta g_{r-} & = & \tilde{h}_{r-}\ r^{-3}\log\left(  r\right)  +\tilde
{f}_{r-}r^{-3}+\cdot\cdot\cdot\\
\Delta g_{++} & = & f_{++}+\cdot\cdot\cdot\\
\Delta g_{+-} & = & f_{+-}+\cdot\cdot\cdot\\
\Delta g_{--} & = & \tilde{h}_{--}\;\log\left(  r\right)  +\tilde{f}%
_{--}+\cdot\cdot\cdot
\end{array}
\label{Asympt metric chiral}%
\end{equation}
which accommodates solutions of the form (\ref{pp-wave-mu-chiral}) having
constant curvature at the asymptotic region. The variation of the charges is
then obtained following the same procedure as explained above for the
asymptotic behavior of the metric in (\ref{Asympt metric chiral}). The only
nonvanishing terms in (\ref{surfaceappendix}) are then given by%
\begin{align*}
\xi_{-}^{\mu}e_{\;\mu}^{a}\delta\beta_{a\phi}  &  =\frac{2}{l}T^{-}%
\delta\tilde{h}_{--}\\
& \\
\xi_{+}^{\mu}e_{\;\mu}^{a}\delta A_{a\phi}  &  =\frac{1}{l}T^{+}\delta f_{++}%
\end{align*}
so that the variation of the charges reads \cite{Henneaux:2009pw}%
\begin{align*}
\delta Q_{\pm}  &  =\frac{4}{l}\int T^{+}\delta f_{++}d\phi\ ,\\
\delta Q_{-}  &  =\frac{2}{l}\int T^{-}\delta\tilde{h}_{--}d\phi\ .
\end{align*}
The charges are then given by
\begin{equation}
Q_{+}=\frac{4}{l}\int T^{+}f_{++}d\phi\ ,\text{ and }Q_{-}=\frac{2}{l}\int
T^{-}\tilde{h}_{--}d\phi\ . \label{Q+Q-Chiral}%
\end{equation}
Note that $Q_{-}[T^{-}]$ does not vanish identically, but rather, the
relaxation term $\tilde{h}_{--}$ does contribute to it. This behavior is
somehow similar to what occurs for scalar fields that saturates the BF bound
\cite{Henneaux:2004zi}.

Note also that the Virasoro generators with both chiralities are non zero at
the chiral point (while one chiral set of them does vanish under the boundary
conditions of \cite{Brown:1986nw}). The asymptotic form of the metric
(\ref{Asympt metric chiral}) agrees with the one in
\cite{Grumiller-Johansson2} and it can be obtained from
(\ref{Asympt relaxed metric mu}) in the limit $\mu l\rightarrow1$. This can be
seen as follows:\ Requiring the curvature to be constant at infinity implies
that the branch with positive chirality in (\ref{Asympt relaxed metric mu})
has to be switched off (i.e., one makes $h_{++}=0$, so that $h_{r +}$ can
be gauged away). Then, in the limit $\mu l\rightarrow1$, since $r^{1-\mu
l}=1+(1-\mu l)\log(r)+\cdot\cdot\cdot$, one recovers
(\ref{Asympt metric chiral}) with \cite{Sezgin-Tanii}%
\begin{equation}%
\begin{array}
[c]{ccc}%
\tilde{h}_{--}=\left(  1-\mu l\right)  h_{--} & ; & \tilde{f}_{--}%
=h_{--}+f_{--}\\
\tilde{h}_{r-}=\left(  1-\mu l\right)  h_{r-} & ; & \tilde{f}_{r-}%
=h_{r-}+f_{r-}%
\end{array}
\label{Redefinition}%
\end{equation}
It is amusing to verify that the charges at the chiral point in Eq.
(\ref{Q+Q-Chiral}) can be obtained from the ones off the chiral point in
(\ref{once}) in the limit $\mu l\rightarrow1$.

\subsection{Integration from infinity}

The field equations can easily be integrated from infinity in the chiral case.
For $\mu l=1$ and the asymptotic conditions given by
(\ref{Asympt metric chiral}), the leading terms of the relevant field
equations reduce to%
\begin{align}
E_{rr}  &  =r^{-4}\left[  -l^{-2}f_{rr}+4f_{+-}\right]  +\cdot\cdot
\cdot\label{ErrLog}\\
E_{r+}  &  =r^{-3}\left[  4\partial_{-}f_{++}\right]  +\cdot\cdot
\cdot\label{Er+Log}\\
E_{r-}  &  =r^{-3}\left[  2\partial_{+}\tilde{h}_{--}\right]  +\cdot\cdot
\cdot\label{Er-Log}%
\end{align}
where the first equation was used to simplify the second and the third ones,
and the remaining field equations are of subleading orders as compared with
the asymptotic behaviour of the metric.

\subsection{Central charges}

We now turn to the computation of the central charges.

In the case of $\mu l=1$, under the action of the Virasoro symmetry
(\ref{Asympt KV}), one obtains the same Eq. (\ref{delta f++ con tutti}), and%
\[
\delta_{\eta}h_{--}=2h_{--}\partial_{-}T^{-}+T^{-}\partial_{-}h_{--}%
+T^{+}\partial_{+}h_{--}\ ,
\]
and the asymptotic field equations (\ref{Er+Log}), (\ref{Er-Log}) read%
\[
\partial_{-}f_{++}=0\text{, and }\partial_{+}h_{--}=0\text{.}%
\]
Note that this time the equations do not impose $\partial_{+}f_{--}=0$ and
furthermore, the transformation rule of $f_{--}$ also differs from the one
found off the chiral point. The variation of the charges (\ref{Q+Q-Chiral})
then reads%
\begin{align*}
\delta_{\eta_{2}}Q_{+}[T_{1}^{+}]  &  =[Q_{+}[T_{1}^{+}],Q_{+}[T_{2}%
^{+}]+Q_{-}[T_{2}^{-}]]\\
&  =\frac{4}{l}\int T_{1}^{+}\ \delta_{\eta_{2}}f_{++}d\phi\ ,\\
&  =\frac{4}{l}\int d\phi\left[  (T_{1}^{+}\partial_{+}T_{2}^{+}-T_{2}%
^{+}\partial_{+}T_{1}^{+})f_{++}-\frac{l^{2}}{2}T_{1}^{+}\left(  \partial
_{+}T_{2}^{+}+\partial_{+}^{3}T_{2}^{+}\right)  \right] \\
&  =Q_{+}\left[  [T_{1}^{+},T_{2}^{+}]\right]  -2l\int d\phi T_{1}^{+}\left(
\partial_{+}T_{2}^{+}+\partial_{+}^{3}T_{2}^{+}\right)  \ ,
\end{align*}
and
\begin{align*}
\delta_{\eta_{2}}Q_{-}[T_{1}^{-}]  &  =[Q_{-}[T_{1}^{-}],Q_{+}[T_{2}%
^{+}]+Q_{-}[T_{2}^{-}]]\\
&  =\frac{2}{l}\int T_{1}^{-}\ \delta_{\eta_{2}}h_{--}d\phi\ ,\\
&  =\frac{2}{l}\int d\phi\ (T_{1}^{-}\partial_{-}T_{2}^{-}-T_{2}^{-}%
\partial_{-}T_{1}^{-})h_{--}\\
&  =Q_{-}\left[  [T_{1}^{-},T_{2}^{-}]\right]  \ .
\end{align*}
Therefore, one finds that both $Q_{+}[T^{+}]$ and $Q_{-}[T^{-}]$ fulfill the
Virasoro algebra with the central charges
\begin{equation}
c_{+}=2\,c\,,\;\;\;\;c_{-}=0\,.
\end{equation}
The central charge $c_{-}$ is vanishes because the inhomogeneous terms
$-l^{2}\left(  \partial_{-}T^{-}+\partial_{-}^{3}T^{-}\right)  /2$ are absent
from $\delta_{\eta}\tilde{h}_{--}$.

The vanishing of the central charge $c_{-}$ is somehow puzzling from the point
of view of conformal field theory. Progress in this direction has been
recently achieved in Refs. \cite{ChiralCFTs}, \cite{LCFTs},
\cite{Andrade-Marolf}, \cite{Becker-Chiral-Supergravity}.

The results of this section have been confirmed following different approaches
in Refs. \cite{Sezgin-Tanii}, \cite{Maloney-Song-Strominger}, and
\cite{Skenderis2} (notice that \cite{Sezgin-Tanii} and \cite{Skenderis2}
consider also the non-chiral point with only one chirality).

\end{document}